\documentstyle{article}

\makeatletter

\def\@aabuffer{}
\def\author #1{\expandafter\def\expandafter\@aabuffer\expandafter
{\@aabuffer \small\rm      #1\relax \par}}
\def\address#1{\expandafter\def\expandafter\@aabuffer\expandafter
{\@aabuffer \small\it #1\relax \par\vspace{1em}}}

\def\maketitle{
\begin{center}
   {\bf \@title \par}
   \vskip 2em                      
   \@aabuffer\relax
\end{center} \par
\gdef\@aabuffer{}
}

\def\abstracts#1{
\begin{center}
{\begin{minipage}{4.2truein}
                 \footnotesize
                 \parindent=0pt #1\par
                 \end{minipage}}\end{center}
                 \vskip 2em \par}

\fussy
\def\section{\@startsection {section}{1}{\z@}{-3.25ex plus -1ex minus
    -.2ex}{1.5ex plus .2ex}{\bf }}
\def\subsection{\@startsection{subsection}{2}{\z@}{-3.25ex plus -1ex minus
    -.2ex}{1.5ex plus .2ex}{\it }}

\def\thefootnote{\alph{footnote}}

\renewcommand{\@makefntext}[1]{\noindent{\@makefnmark}#1} 

\renewenvironment{thebibliography}[1]
	{\begin{list}{\arabic{enumi}.}
	{\usecounter{enumi}\setlength{\parsep}{0pt}
	 \setlength{\itemsep}{0pt}
         \settowidth
	{\labelwidth}{#1.}\sloppy}}{\end{list}}

\newcounter{arabiclistc}

\def\@citex[#1]#2{\if@filesw\immediate\write\@auxout
	{\string\citation{#2}}\fi
\def\@citea{}\@cite{\@for\@citeb:=#2\do
	{\@citea\def\@citea{,}\@ifundefined
	{b@\@citeb}{{\bf ?}\@warning
	{Citation `\@citeb' on page \thepage \space undefined}}
	{\csname b@\@citeb\endcsname}}}{#1}}

\newif\if@cghi
\def\cite{\@cghitrue\@ifnextchar [{\@tempswatrue
	\@citex}{\@tempswafalse\@citex[]}}
\def\citelow{\@cghifalse\@ifnextchar [{\@tempswatrue
	\@citex}{\@tempswafalse\@citex[]}}
\def\@cite#1#2{{$\!^{#1}$\if@tempswa\typeout
	{IJCGA warning: optional citation argument
	ignored: `#2'} \fi}}

\def\baselinestretch{1.0}
\ifx\selectfont\undefined
\@normalsize\else\let\glb@currsize=\relax\selectfont
\fi

\ifx\selectfont\undefined
\def\@singlespacing{%
\def\baselinestretch{1}\ifx\@currsize\normalsize\@normalsize\else\@currsize\fi%
}
\else
\def\@singlespacing{\def\baselinestretch{1}\let\glb@currsize=\relax\selectfont}
\fi

\long\def\@makecaption#1#2{
   \vskip 10pt
   \setbox\@tempboxa\hbox{\footnotesize #1: #2}
 \ifdim \wd\@tempboxa >\hsize \footnotesize #1: #2\par \else \hbox
 to\hsize{\hfil\box\@tempboxa\hfil}
   \fi}

\makeatother

\def\Journal#1#2#3#4{{#1} {\bf #2}, #3 (#4)}

\def\NPB{{\em Nucl. Phys.} B}
\def\PLB{{\em Phys. Lett.}B}

\def\st{\scriptstyle}
\def\sst{\scriptscriptstyle}
\def\tr{{\rm tr\,}}
\def\for{\lower6pt\hbox{$\Big|$}}

\def\be{\begin{equation}}
\def\ee{\end{equation}}
\def\bea{\begin{eqnarray}}
\def\eea{\end{eqnarray}}
\def\nn{\nonumber}

\def\ft#1#2{{\textstyle{{\scriptstyle #1}\over {\scriptstyle #2}}}}
\def\fft#1#2{{#1 \over #2}}

\def\st{\scriptstyle}
\def\sst{\scriptscriptstyle}
\def\gtlt{\mathrel{\raise4.5pt\hbox{\oalign{$\scriptstyle>$\crcr
$\scriptstyle<$}}}}
\def\ffrac#1#2{\leavevmode\kern.1em
\raise.3ex\hbox{\the\scriptfont0 #1}\kern-.1em/\kern-.15em
\lower.2ex\hbox{\the\scriptfont0 #2}}
\def\sffrac#1#2{\leavevmode\kern.1em
\raise.3ex\hbox{\the\scriptscriptfont0 #1}\kern-.1em/\kern-.15em
\lower.2ex\hbox{\the\scriptscriptfont0 #2}}
\def\gtlt{\mathrel{\raise4.5pt\hbox{\oalign{$\scriptstyle>$\crcr
$\scriptstyle<$}}}}

\def\isomorphic{\mathrel{\raise4pt\hbox{\oalign{$\scriptstyle\sim$\crcr
$\scriptstyle=$}}}}
\def\R{\rlap{\hbox{\rm I}}\mkern3mu\hbox{\rm R}}
\def\C{\mkern1mu\raise2.2pt\hbox{$\scriptscriptstyle|$}\mkern-6mu\hbox{\rm C}}

\def\Z{\rlap{\hbox{\sf Z}}\mkern3mu\hbox{\sf Z}}
\def\for{\lower3pt\hbox{$\st|$}}

\def\im{{\rm i}}

\def\dalemb#1#2{{\vbox{\hrule height .#2pt
        \hbox{\vrule width.#2pt height#1pt \kern#1 pt
                \vrule width.#2 pt}
        \hrule height.#2 pt}}}

\begin{document}
\renewcommand{\thefootnote}{\fnsymbol{footnote}}
\title{DOMAIN WALLS AND THE UNIVERSE\footnote{Based upon a talk given at the
Conference on Superfivebranes and Physics in 5+1 Dimensions, Trieste, Italy, 1-3
Apr 1998}}
\author{K.S. STELLE}
\address{The Blackett Laboratory, Imperial College,\\
Prince Consort Road, London SW7 2BZ, UK}

\maketitle\raisebox{3.75cm}[0pt]{Imperial/TP/98--99/24\hskip46mm hep-th/9812086}
\abstracts{$D=11$ supergravity possesses $D=5$ Calabi-Yau compactified solutions
that may be identified with the vacua of the Ho\v{r}ava-Witten orbifold
construction for M--theory/heterotic duality. The simplest of these solutions
naturally involves two 3-brane domain walls, which may be identified with the
orbifold boundary planes; this solution also possesses an unbroken $\Z_2$
symmetry. Consideration of nearby excited solutions, truncated to the zero-mode
and $\Z_2$ invariant sector, yields an effective $D=4$ heterotic theory
displaying chirality and $N=1$, $D=4$ supersymmetry.}
\renewcommand{\thefootnote}{\alph{footnote}}

\section{Introduction}
\label{sec:intro}
One of the striking aspects of the ongoing reformulation of
string theory is the extent to which supergravity effective
field theories can provide important non-perturbative information about the
underlying quantum theory. Various duality relations have been
proposed to hold between the different perturbatively consistent string
theories, and also between these theories and the anticipated quantum precursor
to $D=11$ supergravity, which has been called M--theory. We still lack, however,
a satisfactory microscopic formulation of quantum M--theory. Thus, duality
relations between  M--theory and the perturbative string theories remain
somewhat tentative. Accordingly, it is important to establish just which aspects
of a duality relation are in fact already present at the effective
field-theoretic level, and which truly involve quantum phenomena.

An important case in point is the Ho\v{r}ava-Witten relation between M--theory
and $D=10$ heterotic string theory.\cite{hw1,hw2} The essential idea in this
relation involves compactification of the $D=11$ theory on an orbifold,
causing a sensitivity to anomalies in an otherwise anomaly-free theory.
Requiring cancellation of the anomalies proves to be the essential clue that
reveals the presence of supersymmetric Yang-Mills modes propagating only in
the two orbifold fixed planes that bound the $D=11$ spacetime. In the limit
where these boundary planes lie close together, these $D=10$ zero modes
are described by $D=10$ heterotic string theory.

One aspect of the M--theory/heterotic duality that remains somewhat obscure in
the  Ho\v{r}ava-Witten scenario is whether the appearance of the orbifold can be
viewed as a natural occurrence in M--theory, or whether this is being introduced
from the outside as an extraneous ingredient. Given the absence of an underlying
microscopic formulation of M--theory, this question is hard to answer at the
quantum level. But one can still try to gain what insight one can into this
question at the effective field-theoretic level of $D=11$ supergravity.

Reduction of the $D=10$ theory obtained from M--theory through the orbifold
construction on down to $D=4$ proceeds in a fairly standard fashion by further
compactification on a Calabi-Yau manifold. Searching for a solution to the
resulting $D=4$ theory then reveals~\cite{w} a deformed Calabi-Yau vacuum
solution, found originally at lowest order in a perturbative expansion in
$\kappa^{\fft23}$, where $\kappa$ is the $D=11$ gravitational coupling constant.
The deformation of the Calabi-Yau solution involves, specifically, an evolution
in the volume of the Calabi-Yau manifold as one moves in the bounded orbifold
coordinate, originally taken to be $x^{11}$. 

This behavior underlines another salient aspect of M--theory/heterotic
duality. In modern Kaluza-Klein theory, considerable attention has been
paid to the question of {\em consistency} of the dimensional reduction. By
consistency, one means that solutions to the dimensionally reduced theory are
still perfectly good, albeit rather specific, solutions to the original
higher-dimensional theory. Variation of the Calabi-Yau volume as one moves
across the orbifold indicates clearly that this $11\rightarrow10$ reduction is
not consistent in this technical sense. Now, it is not automatically disastrous
to get involved in a technically inconsistent Kaluza-Klein reduction like this.
But what is required to treat such a case is a careful integrating out, or
averaging, of the non-trivial higher dimensional modes in order to extract the
effective lower dimensional physics. In consistent Kaluza-Klein reductions, by
contrast, the non-trivial higher dimensional modes may simply be set equal to
their vacuum values, which are typically vanishing. Such behavior may in fact be
taken to give an alternative definition of a consistent reduction. For the
present case of an inconsistent orbifold compactification, it seems more
appropriate instead to retain the orbifold dimension within the
lower-dimensional theory, leading thus to a $D=5$ perspective on the
Ho\v{r}ava-Witten construction.

In contrast to the orbifold reduction, the straightforward reduction of
supergravity theories on Calabi-Yau manifolds may be considered to be
``essentially consistent'' in the sense that any corrections induced by
integrating out the non-zero-modes can only be of higher order in derivatives
than the leading order effective field theory.\cite{dfps} The possibility of
higher-derivative corrections to the effective theory presumably also
corresponds to uncertainty about just what the explicit form of the Kaluza-Klein
ansatz for a Calabi-Yau reduction ought to be, since the Calabi-Yau metrics 
are not known explicitly.

Here, we shall review from the above perspective another approach to the
derivation of $D=4$ physics from M--theory compactified on a Calabi-Yau
manifold.\cite{losw1,losw2} Taking a clue from the above discussion that the
physics of the Ho\v{r}ava-Witten construction should really be viewed as five
dimensional, we shall first dimensionally reduce $D=11$ supergravity on a
Calabi-Yau manifold down to $D=5$. Finding the right vacuum solution, which
will extend the solution of Ref.\,\cite{w} to all orders in the gravitational
coupling constant, will then reproduce for us the remaining salient features of
the Ho\v{r}ava-Witten construction: the $\Z_2$ orbifold, chirality and $N=1$,
$D=4$ supersymmetry, but this shall now occur within the natural context of
solutions to $D=11$ supergravity.

\section{Generalized Calabi-Yau reduction of M--theory}
\label{sec:CYred}

Start from the $D=11$ supergravity action, whose bosonic sector is
\be
I_{11}=\int d^{11}x\sqrt{-g}(R-\ft1{48}G_{[4]}^2)+\ft16G_{[4]}\wedge G_{[4]}
\wedge A_{[3]}\ ,\label{d11act}
\ee
where $G_{[4]}=dA_{[3]}$ is the field strength for the 3-form gauge potential
$A_{[3]}$. When making a standard compactification~\cite{CYred} 
${\cal M}_{11}={\cal M}_5\times{\cal X}$ on a Calabi-Yau 3-fold $\cal X$
which is characterized by Hodge numbers $h^{(1,1)}$, $h^{(2,1)}$ and intersection
numbers $d_{ijk}=\int_{\cal X}\omega_i\wedge\omega_j\wedge\omega_k$, one obtains 
$N=1$, $D=5$ supergravity theory coupled to $(h^{(1,1)}-1)$ $D=5$ vector
multiplets, plus one universal hypermultiplet, plus $h^{(2,1)}$ additional
hypermultiplets. Although there are a total of $h^{(1,1)}$ vector fields arising
in the Kaluza-Klein reduction on ${\cal X}$, there are only $(h^{(1,1)}-1)$
vector multiplets because one vector is required in the $N=1$, $D=5$ supergravity
multiplet. 

The universal hypermultiplet contains the modulus field $V=\ft16
d_{ijk}a^ia^ja^k$ corresponding to the volume of the Calabi-Yau manifold ${\cal
X}$, where the $a^i$ are general $h^{(1,1)}$ moduli. The $a^i$ correspond
to deformations of the complex structure: $\omega_{\sst
AB}=a^i\omega_{i{\sst AB}}$, where the $\omega_{i{\sst AB}}$, $i=1,\ldots
h^{(1,1)}$ form a basis of the (1,1) forms on the Calabi-Yau manifold (where
${\st A}=a,\bar a$ run over the six internal coordinate directions). In the form
directly inherited from dimensional reduction, the universal hypermultiplet
also contains the $D=5$ 3-form gauge field $A_{\alpha\beta\gamma}$ and also a
complex scalar $\xi$ obtained from the purely Calabi-Yau components of the
$D=11$ 3-form: $A_{abc}=\ft16\xi\Omega_{abc}$, where
$\Omega_{abc}$ is the harmonic (3,0) form on the Calabi-Yau manifold.

Outside the universal hypermultiplet, there are the $(h^{(1,1)}-1)$ remaining
shape-determining metric moduli that are independent of $V$. These may be denoted
by $b^i=V^{-\fft13}a^i$. These modulus fields are members of the $(h^{(1,1)}-1)$
$D=5$ vector multiplets arising from the dimensional reduction.  For the 
fields $b^i$, one obtains a non-linear sigma model action with metric
$G_{ij}=-\ft12V^{\fft23}{\partial\over\partial a^i}{\partial\over\partial
a^j}\ln V$.

Ordinary dimensional reduction of $D=11$ supergravity on a Calabi-Yau
manifold~\cite{CYred} in this way produces a massless theory of $N=1$, $D=5$
supergravity coupled to $N=1$, $D=5$ supermatter.
The natural subsequent dimensional reduction of this theory down to $D=4$ on a
circle would then give rise to an $N=2$, $D=4$ non-chiral theory. While
interesting, this reduction path does not lead one to realistic-looking $D=4$
physics. So now one may take a clue from another aspect of the known vacuum
solution~\cite{w} to the Ho\v{r}ava-Witten construction: the 4-form field
strength $G_{\sst ABCD}$ takes non-vanishing values in the internal Calabi-Yau
directions.\cite{losw1,losw2} Normally, letting a differential quantity such as
a field strength take non-vanishing values in compactification directions might
be thought to lead to inconsistency of the reduction. This apparent
inconsistency would arise from forcing the underlying gauge potential to have
non-trivial dependence on the internal Calabi-Yau coordinates, in contrast to
the standard procedure of Kaluza-Klein reduction which suppresses all such
internal coordinate dependence. However, in certain cases one may permit
such  dependence without endangering consistency, provided one uses an extension
of the idea of Scherk-Schwarz reduction, known as generalized dimensional
reduction.\cite{genred}

Generalized Kaluza-Klein reduction has a topological reformulation~\cite{llp} as
follows. One may admit dependence on the internal coordinates in a Kaluza-Klein
reduction provided that such dependence is based upon elements of the homology
groups of the internal manifold ${\cal X}$. Specifically, for an (n-1) form gauge
potential $A_{[n-1]}$ in a space with internal coordinates $z^A$ and with
retained lower-dimensional coordinates $x^\mu$, one may make the consistent
reduction ansatz
\be
A_{[n-1]}(x^\mu,z^{\sst A})=\omega_{[n-1]}(z^{\sst A})+A_{[n-1]}(x^\mu)
\label{genredans}
\ee
provided that $\omega_{[n-1]}$ satisfies
\be
d\omega_{[n-1]}=\Omega_{[n]}\in H^n({\cal X},\R)\ . \label{homcon}
\ee
Simple examples of such reductions~\cite{genred} take a circle ${\cal S}^1$ as
the reduction manifold with $\Omega_{[1]}=mdz\in H^1({\cal S}^1,\R)$, in which
case it is an underlying axion field $A_{[0]}$ that acquires the $z$ dependence:
$A_{[0]}(x^\mu,z)=\omega_{[0]}(z)+A_{[0]}(x^\mu)$. This is cohomologically
non-trivial because, although solving $\omega_{[0]}=mz$ is OK locally, this does
not satisfy globally the required ${\cal S}^1$ periodicity condition.

Now let us apply the generalized reduction procedure to the Calabi-Yau reduction
of M-theory. The Pontryagin class of the Calabi-Yau manifold ${\cal X}$ is
characterized by a set of integers $\beta_i$:
\be
\beta_i={-1\over 8\pi^2}\int_{{\cal C}^i}\tr R\wedge R\ ,\label{betai}
\ee
where the 4-cycle ${\cal C}^i$ is related to the harmonic (2,2) form $\nu_i$
dual to the basis (1,1) form $\omega^i$: $\int_{{\cal C}^i}\nu^j=\delta_i^j$,
$\int_{\cal X}\omega_i\wedge\nu^j=\delta_i^j$. Then, in analogy to
(\ref{genredans}), a generalized Kaluza-Klein reduction on the manifold ${\cal
X}$ may be given by requiring the Calabi-Yau components of $G_{[4]}$ to
take cohomologically nontrivial values
\be
G_{\sst ABCD}=\alpha_i\,\nu^i_{\sst ABCD}\ ,\label{Gans}
\ee
where the coefficients $\alpha_i$ are given by rescaling the topological
integers $\beta_i$ by a factor:
$\alpha_i={\pi\over\sqrt2}\left(\kappa\over4\pi\right)^{2/3}\beta_i$. One may
characterize this generalized reduction as turning on $G_{[4]}$--instantons in
the Calabi-Yau dimensions.

The resulting bosonic $D=5$ reduced theory for the supergravity, universal
hypermultiplet and (1,1) modulus fields has the form~\cite{losw2}
\bea
&&I_5  = \label{d5act}\\
&&\ \  -\fft{1}{2\kappa_5^2}\int_{M_5}d^5x\sqrt{-g}\Big[R+
G_{ij}(b)\partial_\alpha b^i \partial^\alpha b^j + G_{ij}(b){\cal
F}_{\alpha\beta}^i {\cal F}^{j\alpha\beta} 
\nn \\
&&\ \ \hskip6cm + \fft{\sqrt{2}}{12}
\epsilon^{\alpha\beta\gamma\delta\epsilon} d_{ijk}{\cal A}_\alpha^i{\cal
F}_{\beta\gamma}^j{\cal F}_{\delta\epsilon}^k\Big]
\nn\\
&&\ \ 
-\fft{1}{2\kappa_5^2}\int_{M_5}d^5x\sqrt{-g}\Big[\fft{1}{2}V^{-2}\partial_\alpha
V\partial^\alpha V+2V^{-1}\partial_\alpha\xi\partial^\alpha\bar\xi
+\fft{1}{24}V^2G_{\alpha\beta\gamma\delta}G^{\alpha\beta\gamma\delta}
\nn\\
&&\ \ \qquad+\fft{\sqrt{2}}{24}\epsilon^{\alpha\beta\gamma\delta\epsilon}
G_{\alpha\beta\gamma\delta}\big(\im(\xi\bar{X}_\epsilon-\bar{\xi}X_\epsilon) +
2\alpha_i{\cal A}_\epsilon^i\big)
+\fft{1}{2}V^{-2}G^{ij}(b)\alpha_i\alpha_j\Big]\ .\nn
\eea

In the reduced action (\ref{d5act}), one may note the following salient features:
\begin{enumerate}
\item[1)] A {\em potential}
$\fft{1}{2}V^{-2}G^{ij}(b)\alpha_i\alpha_j$ has appeared for the moduli.
\item[2)] The Kaluza-Klein vector fields ${\cal A}^i_\alpha$ participate in {\em
gauge couplings} to the universal hypermultiplet.
\item[3)] The gauge coupling linear in $\alpha_i$ in the last line of
(\ref{d5act}) breaks the $x^5\rightarrow -x^5$ parity $\Z_2$
invariance possessed by the original $D=11$ action (\ref{d11act}).
\end{enumerate}

The action (\ref{d5act}) may be cast into a more standard form by performing a
duality transformation on the $D=5$ 4-form field strength
$G_{\alpha\beta\gamma\delta\epsilon}$: introduce a Lagrange multiplier $\sigma$
in order to impose the Bianchi identity for this field strength as an equation of
motion. Then, $G_{\alpha\beta\gamma\delta\epsilon}$ can be taken to be an
independent field and can be eliminated by its now-algebraic field equation. The
result for the universal hypermultiplet $(V,\sigma,\xi,\bar\xi)$ is an action for
an \ffrac{SU(2,1)}{SU(2)$\times$ U(1)} nonlinear sigma model. The gauge coupling
observed in 2) above acts on the ${\rm U}(1)$ factor in the denominator group.
This gauge coupling allows the elimination of the $\sigma$ field, with
the consequent appearance of a mass term for the vector-field combination
$\alpha_i{\cal A}^i_\alpha$. This structure is characteristic of a gauged
coupling of $D=5$ supergravity to a nonlinear sigma model as was found in the
original studies of $D=5$ supergravity.\cite{gst}

\section{Domain walls and the $\Z_2$ orbifold in 4.5 dimensions}
\label{sec:domwall}

Observation 1) above on the structure of the reduced action
(\ref{d5act}) is characteristic of theories obtained by generalized Kaluza-Klein
reduction.\cite{genred} The presence of a potential for the moduli has a
dramatic effect on the vacuum structure of this dimensionally-reduced theory:
flat space, and, more generally, any maximally symmetric spacetime does not
occur as a solution to the resulting equations of motion. Instead, it appears
that the solution displaying the highest degree of symmetry (including
supersymmetry) for such a theory is actually a {\em domain wall}. In the absence
of a maximally-symmetric solution, a system of domain walls would seem to be the
best candidate for a ``vacuum'' solution to the theory. Thus, the theory
(\ref{d5act}) is one in which translation invariance appears to be spontaneously
broken.

Trying a domain-wall ansatz for a solution to the $D=5$ theory (\ref{d5act}),
one posits
\bea
ds_5^2 &=& a(y)^2dx^\mu dx^\nu\eta_{\mu\nu}+b(y)^2dy^2\nn\\
V &=& V(y)\qquad\qquad b^i\ =\ b^i(y)\ ,\label{wallans}
\eea
where $\mu,\nu=0,1,\ldots,3$ will remain as a set of $D=4$ Lorentz coordinates
and $y=x^5$ will play the r\^ole of the coordinate transverse to the domain
wall. Analyzing the resulting equations, one finds~\cite{losw2} a solution with
\be
V(y) = (\ft16 d_{ijk}f^if^jf^k)^2\ ,\quad  a(y) = \tilde k V^{\fft16}\ ,\quad 
b(y) = kV^{\fft23}\ ,\quad  b^i(y) = V^{-\fft16}f^i\ ,\label{domwall}
\ee
where $k,\tilde k$ are integration constants and the functions $f^i(y)$ are
determined implicitly in terms of a set of $h^{(1,1)}$ harmonic functions
$H_i(y)$:
\be
d_{ijk}f^jf^k = H_i\qquad\qquad H_i(y)=2\sqrt2k\alpha_iy + k_i\ ,\label{harmf}
\ee
where the $k_i$ are a further set of $h^{(1,1)}$ integration constants.

The solution (\ref{wallans}--\ref{harmf}) is still not yet satisfactory, for it
leads to singularities at the zeros of the harmonic functions $H_i(y)$.
But this is a standard problem with codimension-one solutions, for which the
harmonic functions are linear, as in (\ref{harmf}). The solution is to replace
the linear dependence on $y$ by dependence on $|y|$, $H_i(y)=2\sqrt2k\alpha_i|y|
+ k_i$. This causes a ``kink'' in the harmonic function at $y=0$, corresponding
to the location of the domain wall. This modification still does not fully cure
the problem of the singularities, however: although analysis of the curvature
shows the metric to go flat as $y\rightarrow\pm\infty$, the scalar field $V(y)$
still blows up in these limits. The cure to this remaining problem is to declare
the $y=x^5$ coordinate to be on a compact ${\cal S}^1$ dimension by identifying
points with $y=\pm\pi\rho$. After this identification, the harmonic function
$H_i(y)$ now has a second kink, corresponding to a second domain wall located at
$y=\pi\rho\leftrightarrow -\pi\rho$.

This construction of the ``vacuum'' resolves the singularity problems of
the na\"{\i}ve ``proto-vacuum'' solution (\ref{domwall}), but, on the other hand,
it does not strictly speaking give a proper solution to the dimensionally reduced
theory (\ref{d5act}). This is because the replacement of $y$ by $|y|$ in the
harmonic function in (\ref{harmf}) amounts to a $\Z_2$ transformation on
(\ref{domwall}) for $y<0$ and, as we have noted in point 3) above, the reduced
action (\ref{d5act}) is not $\Z_2$ invariant. In fact, casual inspection of the
generalized reduction ansatz (\ref{Gans}) shows why this must be the case: one
is setting a $\Z_2$ odd quantity $G_{\sst ABCD}$ to take a non-vanishing $\Z_2$
insensitive fixed value -- thus it is clear that the
$\Z_2$ symmetry must be broken in (\ref{d5act}). What one has achieved in
replacing $y$ by $|y|$ in (\ref{harmf}) is, strictly speaking, to patch together
two non-overlapping regions of the $D=11$ spacetime, with two different 
$G_{[4]}$ Kaluza-Klein ans\"atze related by a $\Z_2$ transformation. One may
write the overall ansatz as
\be
G_{\sst ABCD}=\alpha_i\,\nu^i_{\sst ABCD}\,\epsilon(y)\ ,\label{corrGans}
\ee
where $\epsilon(y)$ is the function that steps between (-1) and 1 as $y$ passes
through zero.

The corrected $G_{[4]}$ ansatz (\ref{corrGans}) actually restores the $\Z_2$
symmetry that was lost in the original ansatz (\ref{Gans}), because a $\Z_2$
transformation combined with a simple rotation of the ${\cal S}^1$ coordinates
by $\pi$ (a special instance of the translations on ${\cal S}^1$ that are
generically broken by the inhomogeneous domain wall solution
(\ref{wallans}--\ref{harmf})) brings the solution back to its original
configuration. Thus, if one were to go beyond the static vacuum solution that we
have been considering, nearby fluctuations could be separated into even and
odd modes under this now-restored $\Z_2$ symmetry.

The most basic mechanism for ensuring consistency of a Kaluza-Klein truncation
or reduction is to make a projection into the invariant sector with respect to
some symmetry. This presents one with a final ``half dimension'' of Kaluza-Klein
truncation that can be made consistently, at the classical level that we have
been discussing, by projection into the even sector under the preserved $\Z_2$
symmetry. This may be mnemonically considered to be reduction to ``4.5
dimensions.'' After such a projection, the retained $D=5$ modes may be
considered to take their values on the spacetime
${\cal M}_4\times{\cal S}^1/\Z_2$. Thus, at the level of the classical modes,
the theory (\ref{d5act}) may be viewed as having spontaneously created the
${\cal S}^1/\Z_2$ orbifold.

\section{Magnetic charge and the cohomology condition}
\label{sec:cohomology}

As a result of the step function in the corrected ansatz (\ref{corrGans}), the
ordinary Bianchi identity for $G_{[4]}$ is no longer valid, but instead one has
\be
(dG)_{5{\sst ABCD}} = {-1\over
4\sqrt2\pi}\left(\kappa\over4\pi\right)^{2/3}(\delta(y)
- \delta(y-\pi\rho))\tr(R\wedge R)_{\sst ABCD}\ ,\label{corrbianchi}
\ee
implying the presence of magnetic charges in the ``vacuum'' solution
(\ref{wallans}--\ref{harmf}). These magnetic charges arise from the
$G_{[4]}$ instantons in the Calabi-Yau directions that have
been turned on by the choice of the topological integers $\beta_i$ (\ref{betai}),
and are occasioned by the $\Z_2$ sign flips as one crosses the domain walls. 

Although the ordinary Bianchi identity for $G_{[4]}$ has been modified to
(\ref{corrbianchi}) in the presence of the domain walls, there is still a
constraint on the fields arising from the global structure of this equation.
Demanding that $(dG)_{[5]}$ be an exact 5-form in the full $D=11$ spacetime, one
obtains a requirement that $\int_{\rm 5-cycle}dG=0$ for integration over an
arbitrary 5-cycle. Picking in particular a 5-cycle ${\cal C}^i\times{\cal S}^1$,
where ${\cal C}^i$ is one of the 4-cycles on the Calabi-Yau manifold, one obtains
the {\em cohomology condition}~\cite{low}
\be
\sum_{{\cal S}^1\ {\rm patches}}\alpha_i = 0\ .\label{cohomcond}
\ee
This is clearly satisfied for the choice of just two proto-vacuum
patches with $\alpha_i^{(2)}=-\alpha_i^{(1)}$, as we have arranged in replacing
$y$ by $|y|$ in the harmonic functions $H_i(y)$ (\ref{harmf}). This condition
also governs the structure of solutions~\cite{low} with more than the
minimal pair of domain walls distributed around the ${\cal S}^1$.

Since we have been at pains to maintain Kaluza-Klein consistency in our
reduction from 11 to 5 dimensions, the $D=5$ double-domain-wall
``vacuum'' solution may also be oxidized back up to $D=11$. In $D=11$, the $D=5$
3-brane is recognized as a 4-dimensional stack of $D=11$ 5-branes wrapped
around 2-cycles of the compactifying Calabi-Yau space, thus effecting a
mixed ``diagonal'' reduction in two dimensions and ``vertical'' reduction in
four dimensions of the compactifying Calabi-Yau manifold.\cite{lps}

\section{Chirality and Supersymmetry}
\label{sec:chiralitysupersymmetry}

In arriving at the above domain-wall solution, we have set the stage for a
compactification to ``4.5'' dimensions, as explained above. The
final half-dimension of truncation is implemented by projecting all $D=5$ fields
into the $\Z_2$ invariant sector. Taking into account the parity oddness of the
$D=11$ $A_{[3]}$ gauge potential as one can see from the action (\ref{d11act}),
this final truncation amounts to imposing conditions like
$g_{\mu\nu}(x^\mu,y)=g_{\mu\nu}(x^\mu,-y)$,
$g_{\mu5}(x^\mu,y)=-g_{\mu5}(x^\mu,-y)$, $A_{\mu\nu\rho}(x^\mu,y)=
-A_{\mu\nu\rho}(x^\mu,-y)$, $\psi_\mu^i(x^\mu,y) =\Gamma_5\psi_\mu(x^\mu,-y)$,
{\it etc.} ($\mu,\nu=0,1,\ldots,3$). Now it becomes clear how the final
reduction to $D=4$ causes chirality to appear. Even though this final step does
not correspond to a consistent Kaluza-Klein reduction and thus one needs to
properly integrate out the non-trivial $D=5$ modes, the chirality implications
of the reduction may still be seen just by suppressing the $y$ dependence in the
$\Z_2$ truncation formulas, yielding chiral fields in $D=4$.

$N=1$, $D=4$ supersymmetry likewise emerges in an expansion of the theory about
the background of the domain-wall ``vacuum'' (\ref{wallans}--\ref{harmf}). This
degree of supersymmetry is reduced with respect to that possessed by the
$D=5$ action (\ref{d5act}) for a reason typical of brane solutions. Such BPS
solutions typically display a reduced degree of supersymmetry with respect to the
vacuum solution of the corresponding theory. In the present case, starting
from a $D=5$ ${\rm SU}(2)$-Majorana spinor parameter $\epsilon^i$ with 8 {\it
a priori} independent components, the domain-wall background has a
surviving supersymmetry requiring $\Gamma_5\epsilon^i=(\tau_3)^i{}_j\epsilon^j$,
cutting the number of independent supersymmetries down to 4. The difference
here with respect to other BPS brane solutions is that there is no other
``vacuum'' solution to the theory (\ref{d5act}) to compare this with, and for
want of another candidate, one might be inclined to consider the
double-domain-wall solution itself as the vacuum. Thus, this solution for the
generalized Calabi-Yau reduction of M--theory presents both a spontaneous
appearance of $D=4$ chirality  and also a spontaneous breakdown of supersymmetry
to the phenomenologically interesting chiral $D=4$, $N=1$ supersymmetry.

\section{Conclusion}
\label{sec:concl}

We have seen that $D=11$ supergravity naturally possesses a solution
compactified down to 4+1 dimensions that reproduces the Ho\v{r}ava-Witten $D=4$
ground state (and in fact which extends this solution to all orders in
$\kappa^{\fft23}$). This solution replaces the orbifold fixed
planes of the Ho\v{r}ava-Witten construction with a pair of standard BPS domain
walls. What has not yet been shown from this Kaluza-Klein perspective is just how
all the fluctuation modes of the domain walls should appear, including in
particular the supersymmetric Yang-Mills multiplets. The outlines of how this
should work can be stated, however. As on the $D=11$ Ho\v{r}ava-Witten
orbifold,\cite{hw1,hw2} the resulting chiral zero-mode theory will be
vulnerable to quantum-level anomalies, and such anomalies will be required to
cancel between the $D=5$ bulk action variations and variations of the $D=3+1$
domain-wall zero-modes, through the mechanism of anomaly inflow.\cite{ch}

This anomaly-cancellation requirement may well allow a rather wide class of
possibilities for the 3-brane zero modes. But at least one example of such a
anomaly-canceling zero-mode set can be given: just reduce the known $D=10$
Ho\v{r}ava-Witten orbifold fixed plane theories~\cite{hw1,hw2} down to $D=5$ on
the Calabi-Yau manifold. If one assumes the standard embedding of the spin
connection into the gauge group, then one ends up~\cite{losw2} with an ${\rm
E}_8$ super Yang-Mills multiplet on one of the $D=3+1$ domain walls and with an
${\rm E}_6$ super Yang-Mills multiplet coupled to chiral supermatter in the {\bf
27} representation on the other, together with the $D=5$ bulk multiplets
discussed above. The general picture that emerges from these
M--theory compactifications is likely to be much richer, however, with many
possibilities for non-standard embeddings~\cite{low} and resulting $D=4$ gauge
groups. Could this be the way in which M--theory finally makes contact with
ordinary $D=4$ physics?

\end{document}